\documentclass[aps,superscriptaddress,showpacs,eqsecnum,secnumarabic,amssymb,amsmath,nofootinbib,tightenlines,showpacs,nobibnotes,aps,prl]{revtex4}
\usepackage{amsfonts}
\usepackage{docs}
\usepackage{bm}
\usepackage{subfigure}
\usepackage[colorlinks=true,linkcolor=blue,citecolor = red]{hyperref}%
\usepackage{graphicx}
\usepackage{dcolumn}
\usepackage{color}
\begin{document}

\title{Influence of Induced Interactions on Superfluid Properties of Quasi-Two Dimensional Dilute Fermi Gases With Spin-Orbit Coupling}

\author{Heron Caldas}\email{hcaldas@ufsj.edu.br}
\affiliation{Departamento de Ci\^{e}ncias Naturais, Universidade Federal de
  S\~ao Jo\~ao Del Rei, 36301-000, S\~ao Jo\~ao Del Rei, MG, Brazil}

\author{ R. L. S. Farias} \email{ricardofarias@ufsj.edu.br}
\affiliation{Departamento de Ci\^{e}ncias Naturais, Universidade Federal de
  S\~ao Jo\~ao Del Rei, 36301-000, S\~ao Jo\~ao Del Rei, MG, Brazil}

\author{Mucio Continentino} \email{mucio@cbpf.br} \affiliation{Centro Brasileiro de Pesquisas Fisicas, Rua Dr. Xavier Sigaud, 150, Urca 
22290-180, Rio de Janeiro, RJ, Brazil}

\date{\today}

\begin{abstract}

We study the effects of induced interactions on the pairing gap, transition temperature and chemical potential of a 
quasi-two dimensional Fermi gas of atoms with spin-orbit coupling. We find that these mean-field parameters are significantly 
modified when induced interactions are taken into account. We also investigate the implications of induced interactions corrections for the BCS-BEC crossover driven by spin-orbit coupling, 
that happens even for small (compared to the Fermi energy) values of the binding energy.

\end{abstract}


\maketitle

\section{Introduction}
The interaction between the spin and orbital degrees of freedom of a particle, known as spin-orbit coupling (SOC), has played important roles in condensed 
matter as well as in atomic and nuclear physics. The main interest to study SOC in ultracold atomic systems (UAS) is the possibility of the observation of (exotic) 
topological manifestations in this context, such the ones that emerge in condensed matter as, for example, topological insulators and technologies such as 
spintronics~\cite{Nayak,Cole}. 

This has greatly motivated the theoretical investigation of the effects of SOC in UAS. For instance, two-dimensional s-wave imbalanced Fermi superfluids with 
SOC may manifest (quantum) topological states of matter~\cite{DasSarma}. Recent studies show that the SOC significantly modifies the Fermi surface and 
produces an enhancement of the low energy density of states~\cite{Vyasanakere, Salasnich}, the triplet pairing and transition temperature~\cite{Z. Q. Yu}, 
the singlet pairing gap~\cite{Chen}, and a suppression of the pair coherence lengths~\cite{B B Huang}.

An effective SOC has been experimentally realized in UAS very recently, where laser beams were applied at rubidium atoms. This allowed the relative strength 
of the interaction to be adjusted by simply tweaking the intensity of the lasers.~\cite{Y. J. Lin1, Y. J. Lin2}.  In this way, SOC in UAS may provide a clean laboratory 
to investigate topological properties of condensed matter systems with the well controlled technique of Feshbach resonance~\cite{Zwierlein, Schunck}.

In this paper we investigate a uniform, homogeneous quasi-two-dimensional (quasi-2D) atomic Fermi gas with Rashba SOC at finite temperatures 
($T$), beyond the mean-field (MF) approximation. The standard MF calculation does not take into account the effects of the medium on the 
two-body interaction. This correction, considered first by Gorkov and Melik-Barkhudarov (GMB)~\cite{GMB61}, usually referred to as induced 
interactions or GMB correction, was found to suppress the MF critical temperatures of a 3D and of a quasi-2D balanced Fermi gases by factors of  
$\approx 2.22$~\cite{GMB61,Pethick00,Baranov2,Yu09} and $\approx 2.72$~\cite{Baranov, Mateus}, respectively. In addition, it has been 
shown that the GMB correction substantially reduces the order parameter in 2D and 3D lattices~\cite{Torma}. The influence of induced 
interactions in a 3D imbalanced Fermi gas has been taken into account in Ref.~\cite{Yu10}, and it was found that the polarization $P_{tc}$ 
and the transition temperature $T_{tc}$ of the tricritical point 
are both reduced from the MF results by a factor of about $2.22$ approaching the experimental value. For a quasi-2D imbalanced Fermi gas, 
$P_{tc}$ and $T_{tc}$ are also reduced by a factor of $\approx 2.72$~\cite{Mateus}, as happens in the spin-balanced configuration. 

We calculate the induced interactions in the presence of SOC. We show that taking into account this generalized induced interactions, the 
enhancement of the superfluid gap parameter found previously in the MF approximation~\cite{Chen}, is significantly decreased both in the region of weak and
strong SOC. We show that the effect of the generalized GMB correction is to decrease $T_c$ by a factor of about $2.72$ 
in weak SOC limit, and by bigger factors in the strong SOC limit. We also investigate the BCS-BEC crossover driven by SOC. We find that in the very low temperature limit, and 
for small two-body binding energy, the generalized GMB correction has almost no effect on the value of the strength of Rashba SOC at which the 
chemical potential changes sign, signaling the crossover from BCS to BEC. However, for higher values of the biding energy, the crossover happens only 
for larger value of the Rashba SOC strength, compared to the MF results for the same binding energies.

The possibility of the Fulde-Ferrel-Larkin-Ovchinnikov (FFLO) state with modulated order parameter~\cite{FFLO} is ignored in this work. As 
in~\cite{GMB61,Pethick00,Baranov,Baranov2,Torma,Yu09}, we consider only pairing between atoms with equal and opposite momenta.

This paper is organized as follows. In Sec.~\ref{sec2}, after introducing the Hamiltonian of a quasi-2D uniform homogenous Fermi gas, we obtain the
finite temperature thermodynamic potential, the gap equation and the number equation for the superfluid phase using the MF approximation. 
In Sec.~\ref{GMB}, we investigate the transition temperature without and with the presence of SOC, at MF and beyond, and map out the phase 
diagram in detail. We conclude in Sec.~\ref{Conc}.

\section{Model Hamiltonian}
\label{sec2}

\begin{figure}[htb]
  \vspace{0.75cm} 
  \includegraphics[width=7.0cm, height=5.5cm]{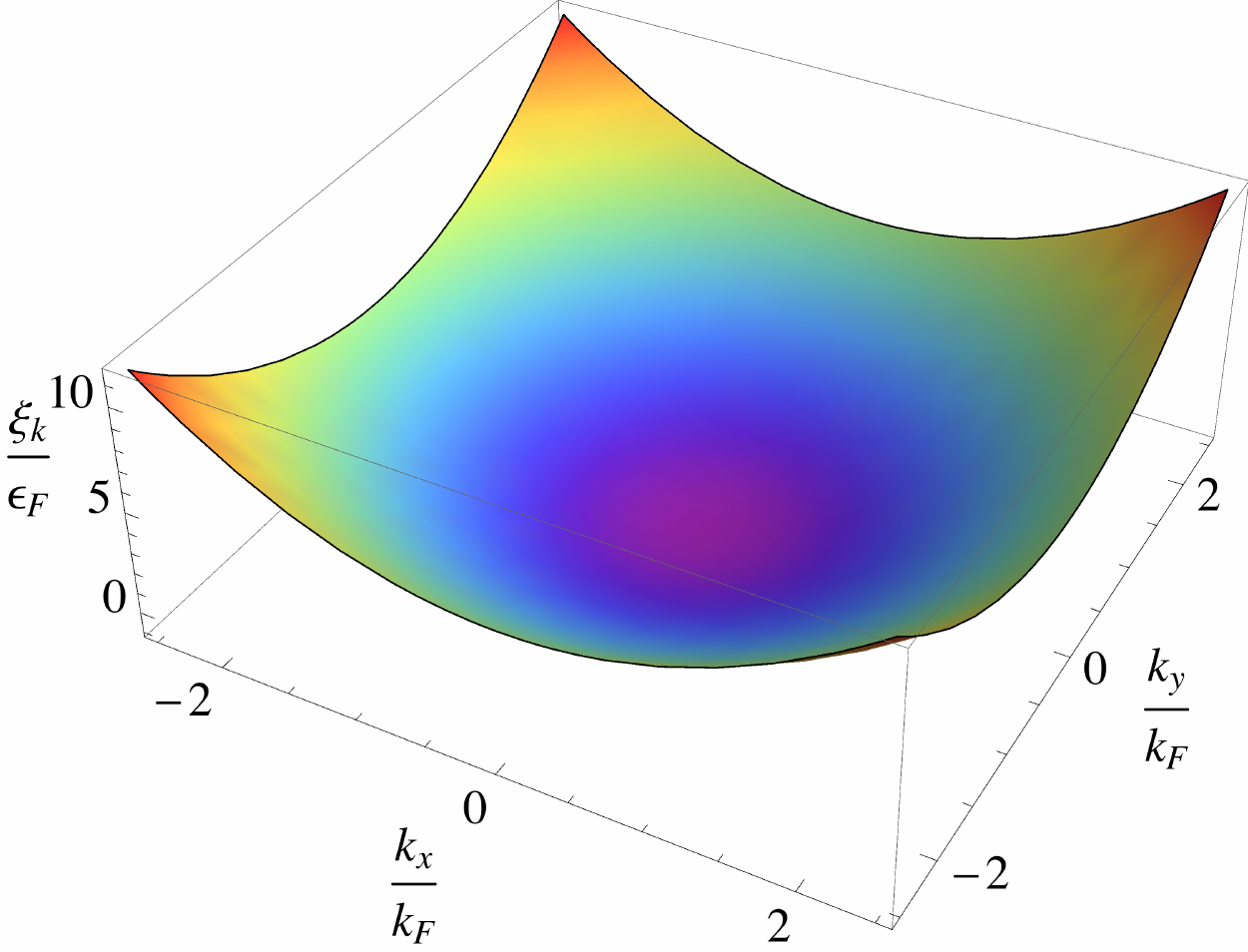}
   \hspace{1.0 cm} \includegraphics[width=7.0cm, height=5.5cm]{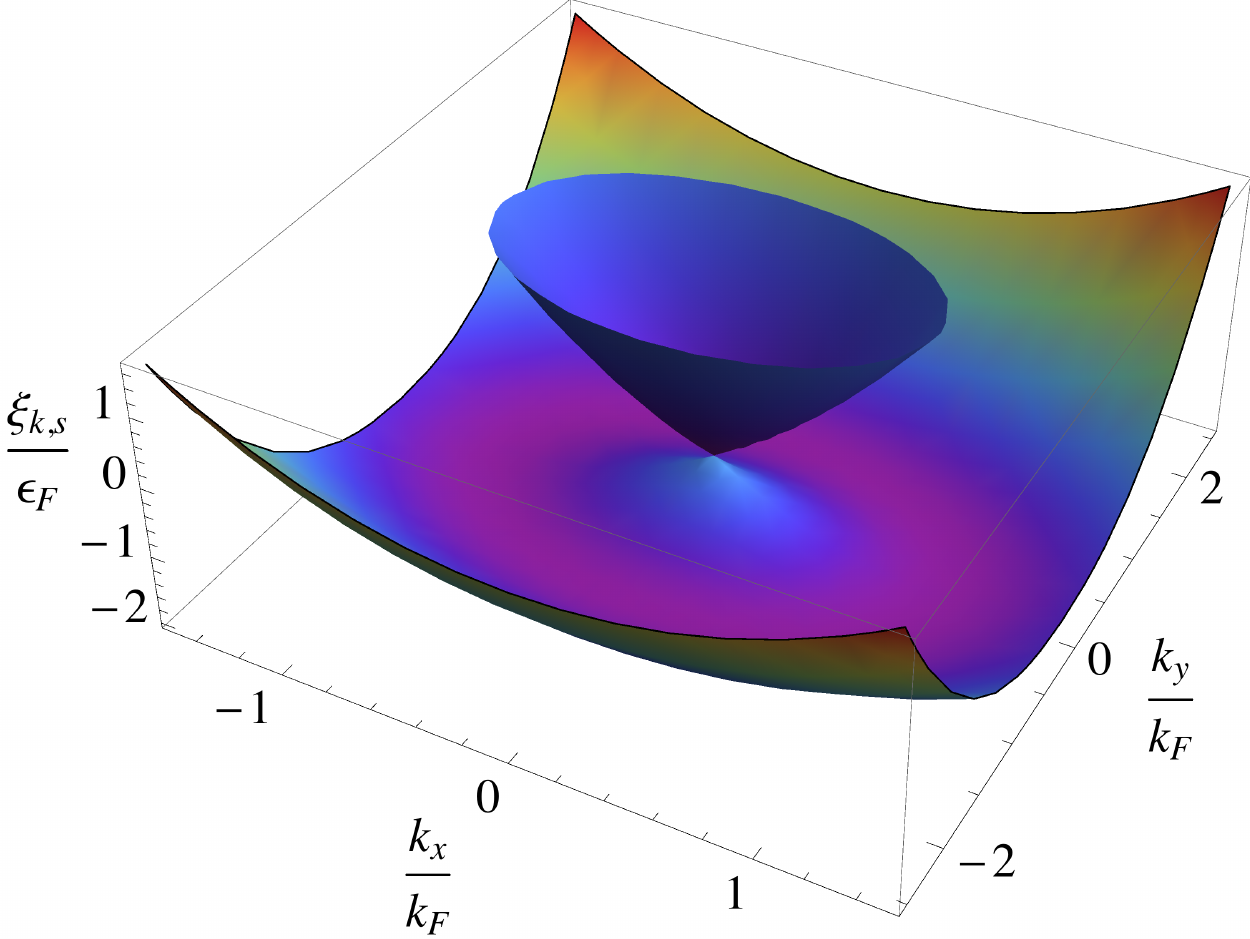}
\caption[]{(Color online) In this figure we show the ordinary single particle spectrum $\tilde{\xi}_{k,s}=\frac{\xi_{k,s}}{\epsilon_F}$ for $\alpha=\frac{\lambda}{v_F}=0$ 
(left panel), and the SOC particle spectrum for $\alpha=1.0$ (right panel).} 
\label{fig1}
\end{figure}

We consider the uniform quasi-2D polarized Fermi gas with SOC, which is
described by the Hamiltonian:
\begin{eqnarray}
H=H_{0}+H_{SO}+H_{int}, \label{1}
\end{eqnarray}
where $H_{0}$ is the kinetic term, $H_{SO}$ is the spin-orbit
interaction, and $H_{int}$ is the s-wave interaction between the two fermionic species, given by
\begin{eqnarray}
&H_{0}&=\sum_{\textbf{k},\sigma} \xi_{\textbf{k},\sigma} c_{\textbf{k},\sigma}^{\dag} c_{\textbf{k},\sigma}, \nonumber\\
&H_{SO}&=\sum_{\textbf{k}} \lambda k \left(e^{-i \varphi_{\textbf{k}}} c_{\textbf{k},\uparrow}^{\dag} c_{\textbf{k},\downarrow} + 
h.c. \right), \nonumber\\
&H_{int}&= g
\sum_{\textbf{k},\textbf{k}'}c_{\textbf{k},\uparrow}^{\dag}
c_{-\textbf{k},\downarrow}^{\dag} c_{-\textbf{k}',\downarrow}
c_{\textbf{k}',\uparrow}, 
\label{2}
\end{eqnarray}
with $\xi_{\textbf{k},\sigma} =\epsilon_k - \mu =\hbar k^{2}/(2m)-\mu$, where $\mu$ is the chemical potential, 
$c_{\textbf{k},\sigma}^{\dag}(c_{\textbf{k},\sigma})$ denotes the
creation(annihilation) operators for a fermion with momentum
$\textbf{k}$ and spin $\sigma=\{ \uparrow, \downarrow \}$, and $\lambda$ is the strength of Rashba spin-orbit coupling,
$\varphi_{\textbf{k}} = \arg(k_{x}+ i k_{y})$, $g<0$ is the (bare) interaction strength (for convenience, we set $\hbar=k_B=1$).
With the transformation,
\begin{eqnarray}
\left(
  \begin{array}{c}
    c_{\textbf{k},\uparrow}\\
    c_{\textbf{k},\downarrow}\\
  \end{array}
\right)=\frac{1}{\sqrt{2}}\left(
  \begin{array}{cc}
    1 & e^{-i \varphi_{\textbf{k}}}\\
    e^{i \varphi_{\textbf{k}}} & -1\\
  \end{array}
\right) \left(
  \begin{array}{c}
    a_{\textbf{k},+}\\
    a_{\textbf{k},-}\\
  \end{array}
\right), \label{4}
\end{eqnarray}
Eq.~(\ref{2}) becomes

\begin{eqnarray}
H_{0}+H_{SO} &=& \sum_{\textbf{k}, s=\pm} \xi_{\textbf{k},s}
a_{\textbf{k},s}^{\dag} a_{\textbf{k},s}, 
\nonumber\\
H_{int} &=& -\frac{\mid \Delta \mid^{2}}{g}  -\frac{\Delta}{2} \sum_{\textbf{k},s=\pm} \left(e^{i s \varphi_{\textbf{k}}} a_{\textbf{k},s}^{\dag}
a_{-\textbf{k},s}^{\dag}  + h.c. \right),
\label{5}
\end{eqnarray}
where the attractive term has been treated using a BCS decoupling. In Eq.~(\ref{5}) $a^{\dag}_{\textbf{k},\pm}(a_{\textbf{k},\pm})$ is 
the creation(annihilation) operator for the state with helicity $(\pm)$, $\xi_{\textbf{k},\pm}=\xi_{\textbf{k}} \pm  \lambda k$ with 
$\xi_{\textbf{k}}=\epsilon_{\textbf{k}}-\mu$, and $\Delta= -g \sum_{\textbf{k}} <c_{-\textbf{k},\downarrow} c_{\textbf{k},\uparrow}>$ 
is the energy gap. In {}Fig.~{\ref{fig1}} we show the single-particle dispersion relations $\xi_{k,\pm}$.

The Hamiltonian in Eq.~(\ref{2}) can be rewritten in the helicity basis~\cite{Alicea}
$\Psi_{{\bf k}} = (a_{\textbf{k},+}, a_{\textbf{k},-},
a_{-\textbf{k},+}^{\dag}, a_{-\textbf{k},-}^{\dag})^{T}$ as:
\begin{eqnarray}
H=\frac{1}{2}\sum_{{\bf k}}\Psi_{{\bf k}}^{\dag}\mathcal{H}({\bf k})\Psi_{{\bf k}}
 + \sum_{\textbf{k}}\xi_{\textbf{k}} - \frac{\mid \Delta \mid^{2}}{g}, \label{6}
\end{eqnarray}
with
\begin{eqnarray}
\mathcal{H}({\bf k})=\left(
  \begin{array}{cccc}
    \xi_{\textbf{k},+} & 0  & -\Delta e^{i \varphi_{\textbf{k}}} &  0  \\
 0 & \xi_{\textbf{k},-}  & 0 & \Delta e^{-i \varphi_{\textbf{k}}} \\
   - \Delta e^{-i \varphi_{\textbf{k}}} & 0 & -\xi_{\textbf{k},+} &  0 \\
    0 & \Delta e^{i \varphi_{\textbf{k}}} & 0 &  -\xi_{\textbf{k},-}  \\
  \end{array}
\right). \label{7}
\end{eqnarray}
This Hamiltonian can be diagonalized as
\begin{eqnarray}
H=\sum_{\textbf{k},s=\pm}
E_{\textbf{k},s}\alpha_{\textbf{k},s}^{\dag}\alpha_{\textbf{k},s}
+ \frac{1}{2} \sum_{\textbf{k},s=\pm}
(\xi_{\textbf{k}}-E_{\textbf{k},s}) - \frac{\mid \Delta
\mid^{2}}{g}, \label{8}
\end{eqnarray}
where, $\alpha_{\textbf{k},\pm}^{\dag}(\alpha_{\textbf{k},\pm})$
is the creation(annihilation) operator for the quasiparticles with
excitation spectra $E_{\textbf{k},\pm}=\sqrt{\xi_{\textbf{k}, \pm}^{2} + \mid \Delta \mid^{2} }$.

\begin{figure}[htb]
  \vspace{0.75cm} 
  \includegraphics[width=10cm, height=6.5cm]{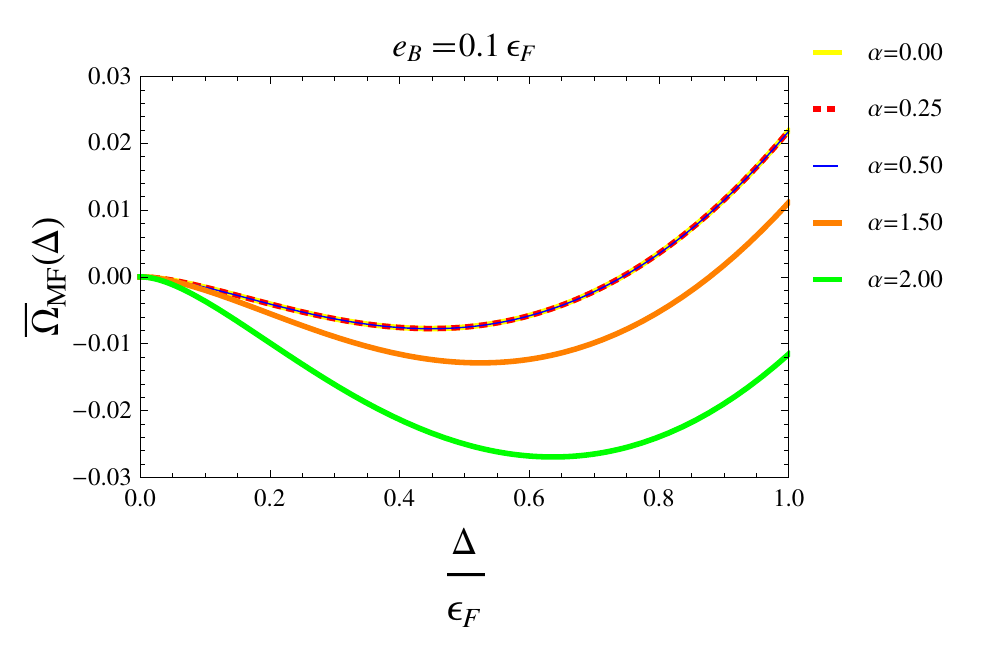}
\caption[]{(Color online) Dimensionless thermodynamic potential $\bar{\Omega}\equiv\frac{\Omega\left(\Delta\right)-\Omega\left(0\right)}{k_F^2\epsilon_F}$ in MF approximation for different values of $\alpha=\frac{\lambda}{v_F}$. The parameters used are $\mu=\epsilon_F$ and $T=0.01\epsilon_F$.}
\label{fig2}
\end{figure}

It is straightforward to write down the grand thermodynamic potential $\Omega = -  \text{Tr}\ln[e^{-\beta H}]$, where $\beta=1/(k_{B} T)$, 
at finite temperatures,

\begin{eqnarray}
\Omega =
\frac{1}{2}\sum_{\textbf{k},s=\pm} \left[\xi_{\textbf{k}}-E_{\textbf{k},s} -\frac{2}{\beta} \ln(1+e^{-\beta E_{\textbf{k},s}}) \right] - \frac{\mid \Delta \mid^{2}}{g}. 
\label{9}
\end{eqnarray}
To regulate the ultraviolet divergence associated with the zero temperature term in Eq.~(\ref{9}), we introduce the 2D bound-state equation~\cite{Randeria}

\begin{equation}
\label{reg} 
-\frac{1}{g}= \int \frac{d^2 k}{(2 \pi)^2} \frac{1}{2\epsilon_k+|\epsilon_B|},
\end{equation}
where $\epsilon_B$ is the 2D two-body binding energy. In addition to the regularization of the ultraviolet divergence present in Eq.~(\ref{9}), 
and consequently in the gap equation below, this equation relates the strength $g$ of the contact interaction with $\epsilon_B$, which is more 
physically relevant, as will be clear now. In order to make contact with current experiments, it is convenient to relate $\epsilon_B$ to the three 
dimensional scattering length $a_s$. In Fig.~\ref{fig2} we show the renormalized (dimensionless) MF thermodynamic 
potential $\bar{\Omega}\equiv\frac{\Omega\left(\Delta\right)-\Omega\left(0\right)}{k_F^2\epsilon_F}$ for different values of the dimensionless constant $\alpha = \frac{ \lambda}{v_F}$. As one can see, 
the minimum $\Delta_0(\alpha)$ increases with increasing $\alpha$, as numerically and analytically shown for $T=0$ in Ref.~\cite{Chen}.

The pairing gap should be self-consistently determined with the chemical potential. This is done by minimizing the thermodynamic potential, i.e. taking $\partial \Omega / \partial \Delta=0$ 
and using the particle number equation $n=-\partial \Omega / \partial \mu$. These yield, respectively, 

\begin{equation}
-\frac{1}{g} = \sum_{k,s}  \frac{1}{2 E_{\textbf{k},s} } \left(\frac{1}{2}- f_k(E_{\textbf{k},s} )\right) ,
\label{gapT}
\end{equation}
where $f_k$ is the Fermi distribution function $f_k(E_{\textbf{k},s} )=1/(e^{\beta E_{\textbf{k},s} }+1)$, and

\begin{eqnarray}
n &=& \sum_{\textbf{k},s=\pm} \left[\frac{1}{2}-\left(\frac{1}{2}- f_k(E_{\textbf{k},s} )\right) 
\frac{\xi_{\textbf{k},s}}{E_{\textbf{k},s}} \right].
\label{NumbEq}
\end{eqnarray}
 We also define the Fermi momentum using $n=k_F^2/2\pi$, with density $n$ (which will be kept fixed throughout our calculation),  so that the Fermi 
velocity is $v_F=k_F/m$.

\section{The Transition Temperature Beyond Mean-Field}
\label{GMB}

Now we construct the phase diagram of the model at finite temperatures beyond the MF approximation, taking into account the GMB correction.

\subsection{Induced interaction in a Fermi gas}
\label{Ind}

\begin{figure}
\includegraphics[width=6.5cm, height=3.0cm]{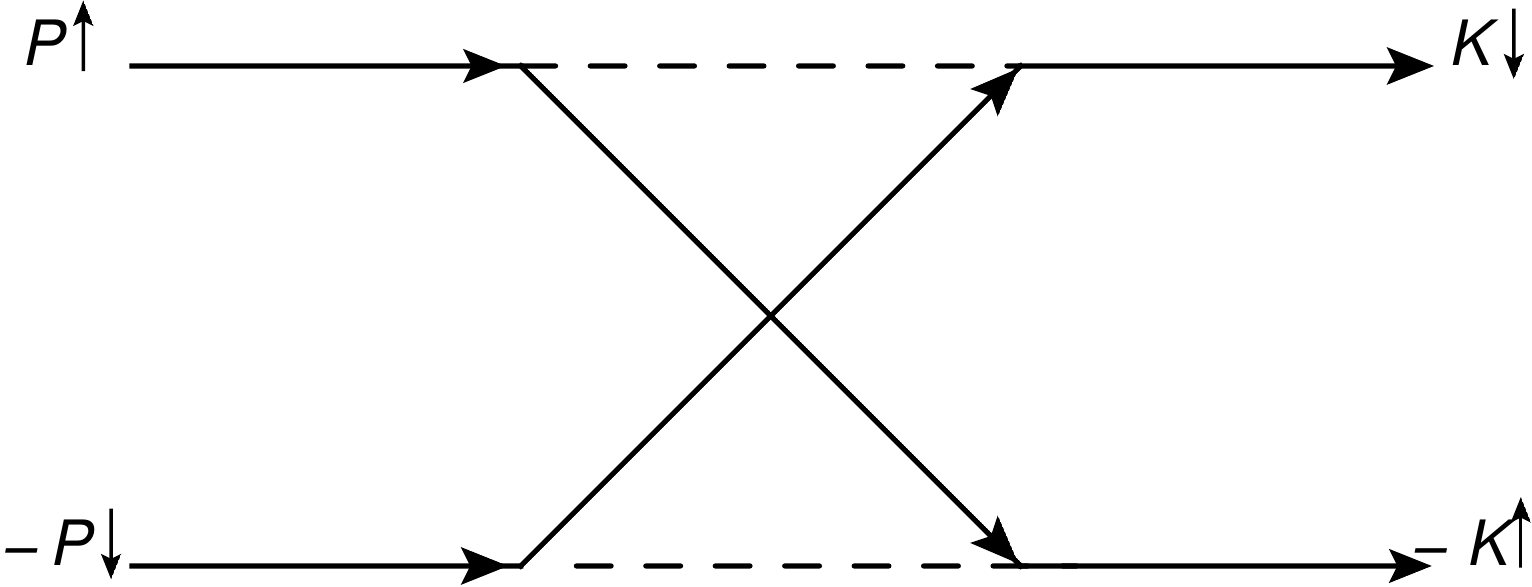}
\caption{The lowest-order diagram representing the induced interaction $U_{\mathrm{ind}}( p_1, p_4)$. Arrowed and dashed lines describe fermionic 
propagators and the coupling $g$ between the fermionic atoms.} 
\label{fig3}
\end{figure}

The induced interaction was obtained originally by GMB in the BCS limit by second-order perturbation theory~\cite{Pethick00}. For a scattering process with 
$p_1+p_2\rightarrow p_3+p_4$, the induced interaction for the diagram in Fig.~\ref{fig3} is expressed as

\begin{equation}
U_{\mathrm{ind}}( p_1, p_4)= -g^2\, \chi_{ph}(p_1-p_4),
\end{equation}
where $p_{i}=({\bf k}_{i}, \omega_{l_i})$ is a vector in the space of wave-vector ${\bf k}$ and fermion Matsubara frequency $\omega_{l}=(2l+1)\pi/(\beta)$. 
Including the induced interaction, the effective pairing interaction between atoms with different spins is given by

\begin{eqnarray}
U_{eff}( p_1, p_4)\equiv U_{eff}&=&
g +U_{\mathrm{ind}}( p_1, p_4).
\label{induce}
\end{eqnarray}
The polarization function $\chi_{ph}(p')$ is given by

\begin{eqnarray}
\chi_{ph}(p')&=& \frac{1}{ \hbar^2 \beta {\rm V}}\sum_p
\mathcal{G}_{0 b}(p)\mathcal{G}_{0 a}(p+p')\\
\nonumber
&=&\frac{\rm{d}^2{\bf k}}{(2\pi)^2} \frac{f_{{\bf k}}^- -f_{{\bf k}+{\bf q}}^+}{ i
\Omega_l+\xi_{{\bf k}}^+ -\xi_{{\bf k}+{\bf q}}^-},
\label{chi0}
\end{eqnarray}
where $p'=(\bf{q}, \Omega_{l})$, $\Omega_{l}=2l\pi/\beta$ is the Matsubara frequency of a boson. The Matsubara Green's function of a 
non-interacting Fermi gas is given by $\mathcal{G}_{0\pm}(p)=1/(i\omega_l-\xi_{\textbf{k}}^{\pm}) $.

In the $\lambda \to 0$ limit the real static polarization function reads

\begin{eqnarray}
\label{chiph-9}
\chi_{ph}(q,\lambda=0) \equiv \chi(q) &=&-N(0) \left[ 1- \sqrt{1-\left( \frac{2 k_F}{q}\right)^2} \right] ~{\rm for}~ q> 2 k_F,  \\
\nonumber
&=& -N(0)~{\rm for}~ q \leq 2 k_F,
\end{eqnarray}
where $N (0) = \frac{m}{2 \pi}$ is the 2D density of states at the Fermi level, and $k_F=\sqrt{2m\mu}$ is the Fermi momentum. The 
induced interaction $U_{ind}(q)$ is then given by~\cite{Pethick00}

\begin{eqnarray}
\label{chiph-9-1}
U_{ind}(q)&=& g^2 N(0) L(x),
\end{eqnarray}
where $L(x)=\left[ 1- \Theta(x-1)\sqrt{1-\left( \frac{1}{x}\right)^2} \right]$. $\Theta(x-1)$ is the step function, and $x \equiv q/2k_F$. In the 
scattering process $\vec{k}_1+\vec{k}_2=\vec{k}_3+\vec{k}_3$, and $q$ is equal to the magnitude of $\vec{k}_1+\vec{k}_3$, so 
$q=\sqrt{(\vec{k}_1+\vec{k}_3).(\vec{k}_1+\vec{k}_3)}=\sqrt{\vec{k}_1^2+\vec{k}_3^2+ 2\vec{k}_1. \vec{k}_3}=\sqrt{\vec{k}_1^2+
\vec{k}_3^2+ 2|\vec{k}_1|| \vec{k}_3|\cos\theta}$. Since both particles are at the Fermi surface, $|\vec{k}_1|=|\vec{k}_3|=k_F$, thus, 
$q=k_F\sqrt{2(1+\cos \theta)}$ and $x=\frac{q}{2k_F}=\frac{\sqrt{2(1+\cos \theta)}}{2}$. Equation~(\ref{chiph-9}) shows that 
the momentum dependence of the critical temperature (i.e. of the order parameter) will appear only due to the second-order term that 
contains many-body contributions to the inter-particle interaction $U_{ind}$. The latter is a function of $q=|\vec{k}_1+\vec{k}_3|$ and 
rapidly decays for $q>2k_F$. For $q<2k_F$ the quantity $U_{ind}$ is a constant.

The {\it s}-wave part of the effective interaction is approximated by averaging the polarization function $U_{ind}(q)$, over the angle 
$\theta$~\cite{GMB61,Pethick00,Yu09,Yu10}:

\begin{equation}
\langle U_{ind}(q)\rangle = \frac{1}{2 \pi} \int_{0}^{2 \pi} d  \theta ~U_{ind}(q) \equiv \bar U_{ind}.
\end{equation}
The fact that both scattering particles are at the Fermi surface resulted in a form for $q$ as $q=k_F\sqrt{2(1+\cos \theta)}$. This means 
that the maximum value for $q$ is obtained for $\theta=0$, which implies $L(x)=1$, yielding

\begin{equation}
\bar U_{ind}=g^2 N(0),
\end{equation}
which gives the reduction of $T_c$ of a quasi-2D Fermi gas \cite{Baranov,Mateus} as

\begin{equation}
\label{TGMB}
T_c = \frac{T_{c0}}{e} \approx \frac{T_{c0}}{2.72},
\end{equation}
where $T_{c0}$ is the MF result. Note that Eq.~(\ref{TGMB}) predicts  the critical temperature to be by a factor of $(1/e)$ smaller than a 
simple MF BCS calculation. This clearly means that the attractive interaction between particles becomes weaker due to the polarization of 
the medium that we have taken into account.

\begin{figure}
\includegraphics[width=8.0cm, height=6.5cm]{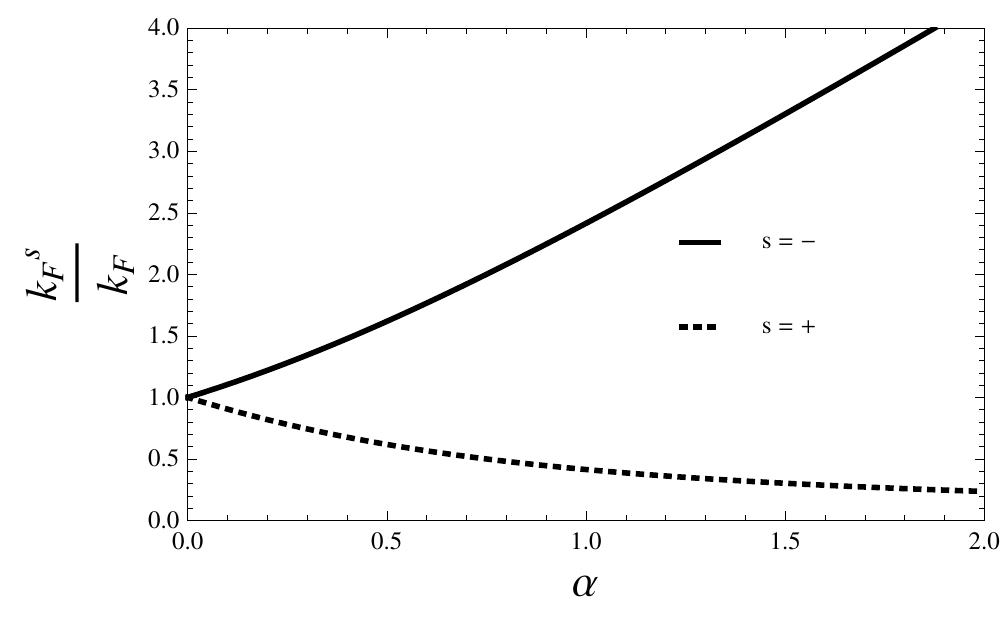}
\caption{Ratio $k_F^{\pm}/k_F$ as a function of $\alpha$. The top curve is $k_F^{-}/k_F$ while the bottom curve is $k_F^{+}/k_F$.} 
\label{fig4}
\end{figure}

Now we consider the static polarization $\chi_{ph}$ in the presence of SOC. With nonvanishing $\lambda$, the calculation of the function 
$\chi_{ph}(q,\lambda)$ is straightforward,

\begin{eqnarray}
\label{chiph-5}
\chi_{ph}(q,\lambda) \equiv \chi(q,\lambda)
&=& - \frac{m}{2 \pi q} \int_0^{\infty} dk~k \left[  \frac{f_{k}^{-}}{\sqrt{\left(\frac{q^2+4 \lambda m k}{2q} \right)^2-k^2}}
+  \frac{f_{k}^{+}}{\sqrt{\left(\frac{q^2-4 \lambda m k}{2q} \right)^2-k^2}}   \right].
\end{eqnarray}
In the zero temperature limit, $ f_{k}^{\pm} \to  k_F^{\pm}$, where the  Fermi momenta modified by the SOC are given by 

\begin{equation}
\label{fm}
k_F^{\pm}=\mp m\lambda + \sqrt{\lambda^2 m^2+ 2m\mu}.
\end{equation}
Fig.~\ref{fig4}  shows the Fermi momenta modified by the SOC $k_F^{\pm}$ normalized by $k_F$, as a function of $\alpha$.

For convenience we define,

\begin{eqnarray}
\label{chiph-6}
\chi(q,\lambda) &\equiv& \chi(q)^{-} + \chi(q)^{+}\\
\nonumber
&=&- \frac{N(0)}{q} \left[ \int_0^{k_F^{-}} dk~k    \frac{1}{\sqrt{\left(\frac{q^2+4 \lambda m k}{2q} \right)^2-k^2}}+  
 \int_0^{k_F^{+}} dk~k  \frac{1}{\sqrt{\left(\frac{q^2-4 \lambda m k}{2q} \right)^2-k^2}}    \right].
\end{eqnarray}
These integrals may be cast in the form

\begin{eqnarray}
\label{chiph-6-1}
\chi(q,\lambda) &=&- \frac{N(0)}{q} \left[ \int_0^{k_F^{-}} dk~k    \frac{1}{\sqrt{ak^2+bk+c}}+   \int_0^{k_F^{+}} dk~k  
\frac{1}{\sqrt{ak^2-bk+c}}    \right],
\end{eqnarray}
where $a=\left(\frac{b}{q} \right)^2-1$, $b=2 \lambda m$, and $c=\left(\frac{q}{2}\right)^2$. The integration in $k$ gives,

\begin{eqnarray}
\label{chiph-6-2}
\int dk~    \frac{k}{\sqrt{ak^2 \pm bk+c}}=\frac{1}{a}\sqrt{ak^2 \pm bk+c} \mp \frac{b}{2 a^{3/2}} \ln[\pm b+2ak+2\sqrt{a}
\sqrt{ak^2 \pm bk+c}] \equiv F^{\mp}(k),
\end{eqnarray}
from which we define $\chi(q,\lambda) =- \frac{N(0)}{q} F^{\mp}(k)=-N(0)F^{\mp}(q,k)$. We adopt the strategy of expanding 
$F^{\mp}(q,k)$ in the relevant limits $\lambda \ll v_F$~\cite{Zeng,Gong}, and $\lambda  \gg v_F$~\cite{Zeng}.
\newline
{\bf 1.} Weak SOC: $\lambda \ll v_F$.

Expanding $F^{\mp}(q,k)$ to first non-vanishing order of $\lambda$, we obtain the real $\chi(q,\lambda)$,

\begin{eqnarray}
\label{chiph-exp1}
\chi(q,\lambda) &=&- N(0) \left\{ \overbrace{1-\sqrt{1-\left(\frac{2 k_F}{q}\right)^2}}^{F(q,\lambda=0)} -4\left[ \frac{q^2- 2k_F^2 -2q 
\sqrt{q^2-(2 k_F)^2}}{q \sqrt{q^2-(2 k_F)^2}}  \right] \frac{m^2 \lambda^2}{q^2} \right\},
\end{eqnarray}
for $q > 2k_F$, and

\begin{eqnarray}
\label{chiph-exp2}
\chi(q,\lambda) &=&- N(0) \left[ 1+ 8 \frac{m^2 \lambda^2}{q^2}  \right],
\end{eqnarray}
for $0<q\leq 2k_F$.
\newline
It is interesting to note that differently from the case without SOC,  $\chi(q,\lambda)$ is now $m$ dependent. The induced interaction reads

\begin{eqnarray}
U_{ind}(q,\lambda)&=& g^2 N(0) F(q,\lambda),
\end{eqnarray}
where we have defined $\chi(q,\lambda) =- N(0) F(q,\lambda)$. Since $q=k_F\sqrt{2(1+\cos \theta)}$, its maximum value is $2 k_F$. Thus, the 
averaged $U_{ind}(q,\lambda)$ will be given by

\begin{equation}
\bar U_{ind}(\lambda) = \frac{1}{2 \pi} \int_{0}^{2 \pi} d \theta ~U_{ind}(q,\lambda)= g^2 \frac{N(0)}{2 \pi} \int_{0}^{2 \pi} 
d \theta ~F(q,\lambda)= g^2 N(0) F(\alpha),
\end{equation}
with

\begin{eqnarray}
\label{fint}
F(\alpha)&=&\frac{1}{2 \pi} \int_{0}^{2 \pi} d \theta ~\left[ 1+ \frac{4 }{1+\cos \theta} \alpha^2 \right]= 1 + 4 \alpha^2 \frac{1}{2 \pi} 
\int_{0}^{2 \pi} d \theta ~ \frac{1 }{1+\cos \theta}\\
\nonumber
&=& 1 +\frac{ 4 \alpha^2}{2 \pi} \tan (\theta/2)~ |_0^{2 \pi}= 1.
\end{eqnarray}
Then, 
$\bar U_{ind}(\lambda) = g^2 N(0)$ which is independent of $\lambda$ to this order, since we just have seen that the angular part of $F(\alpha)$ in 
order $\alpha^2$ averages to zero.
\newline
{\bf 2.} Strong SOC: $\lambda \gg v_F$.

\begin{figure}
\includegraphics[width=8.0cm, height=6.5cm]{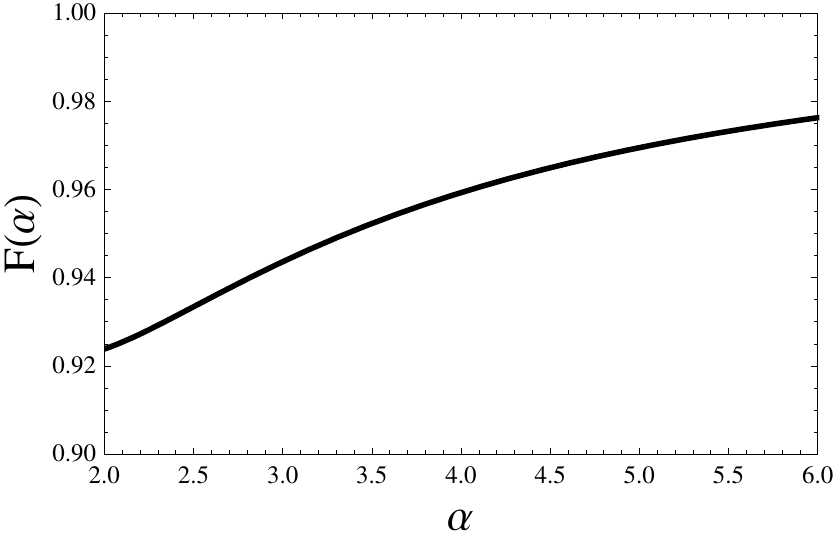}
\caption{Behavior of the GMB correction as a function of (large) $\alpha$.} 
\label{fig5}
\end{figure}

Now we go to the opposite limit, of strong SOC. In this case $F^{\mp}(k,q)$ is greatly simplified,

\begin{equation}
F^{\mp}(k,q)=\frac{k}{b}  \pm \frac{q^2}{2 b^2} \mp \frac{q^2}{2 b^2} \ln \left(\frac{4 b^2}{q^2}k \pm 2b \right),
\end{equation}
from which we obtain

\begin{eqnarray}
\label{chiph-exp3}
\chi(q,\lambda) &=&- N(0) \left[ \sqrt{1+\left(\frac{2 k_F}{b} \right)^2} + \frac{q^2}{2 b^2} \ln \left( \frac{q^2-2b k_F^+}{q^2+2b k_F^-} 
\right) \right].
\end{eqnarray}
Substituting the expressions of $q$, $k_F^+$ and $k_F^-$ in the above equation, yields

\begin{eqnarray}
\label{chiph-exp4}
\chi(q,\lambda) &=&- N(0) \left[ \sqrt{1+\left(\frac{2 k_F}{b} \right)^2} + \frac{k_F^2 }{ b^2} \gamma \ln \left( \frac{k_F^2 2 \gamma +b^2 
- b^2 \sqrt{1+\left(\frac{2 k_F}{b} \right)^2} }
{k_F^2 2 \gamma +b^2 + b^2 \sqrt{1+\left(\frac{2 k_F}{b} \right)^2}} \right) \right],
\end{eqnarray}
where $\gamma \equiv 1+\cos \theta$. After expanding the square roots we get,

\begin{eqnarray}
\label{chiph-exp5}
\chi(q,\lambda) &\approx&- N(0) \left[ 1+ \frac{1}{2} \frac{1}{\alpha^2} +\frac{1}{4} \frac{1}{\alpha^2}  (1+\cos \theta) \ln \left( \frac{\cos \theta }
{\cos \theta + 2 + 4 \alpha^2} \right) \right].
\end{eqnarray}
Expanding now the log term for large $\alpha$, we finally obtain

\begin{eqnarray}
\label{chiph-exp6}
\chi(q,\lambda) &\approx&- N(0) \left[ 1+ \frac{1}{2} \frac{1}{\alpha^2} +\frac{1}{4} \frac{1}{\alpha^2}  (1+\cos \theta) [\ln \left( \cos \theta \right)
- \ln \left(4 \alpha^2 \right)] \right] + {\cal O}\left(\frac{1}{\alpha^4}\right).
\end{eqnarray}
Averaging in $\theta$, 

\begin{eqnarray}
\label{chiph-exp7}
F(\alpha) = &1&+ \frac{1}{2} \frac{1}{\alpha^2} - \frac{1}{4} \frac{1}{\alpha^2} \ln \left(4 \alpha^2 \right)  +\frac{1}{4} \frac{1}{\alpha^2} 
\frac{1}{2\pi} \int_0^{2 \pi} d\theta~ \Theta(\cos \theta) (1+\cos \theta) \ln \left( \cos \theta \right)\\
\nonumber
&1&+ \frac{1}{2} \frac{1}{\alpha^2} - \frac{1}{4} \frac{1}{\alpha^2} \ln \left(4 \alpha^2 \right)  +\frac{1}{4} \frac{1}{\alpha^2} \ln(0.64),
\end{eqnarray}
we are left with the real induced interaction $\bar U_{ind}(\lambda) = g^2 N(0) F(\alpha)$, where

\begin{equation}
F(\alpha) = 1+ \frac{1}{2} \frac{1}{\alpha^2} -\frac{1}{4} \frac{1}{\alpha^2} \ln\left( 6.25 \alpha^2 \right).
\label{Falpha}
\end{equation}
In Fig.~\ref{fig5} we show the behavior of $F(\alpha)$ for large values of $\alpha$. As can be seen the correction approaches $1$ as $\alpha$ 
increases.

\subsection{GMB Corrected Thermodynamic Potential}

\begin{figure}[htb]
  \vspace{0.75cm} 
  \includegraphics[width=10.0cm, height=6.5cm]{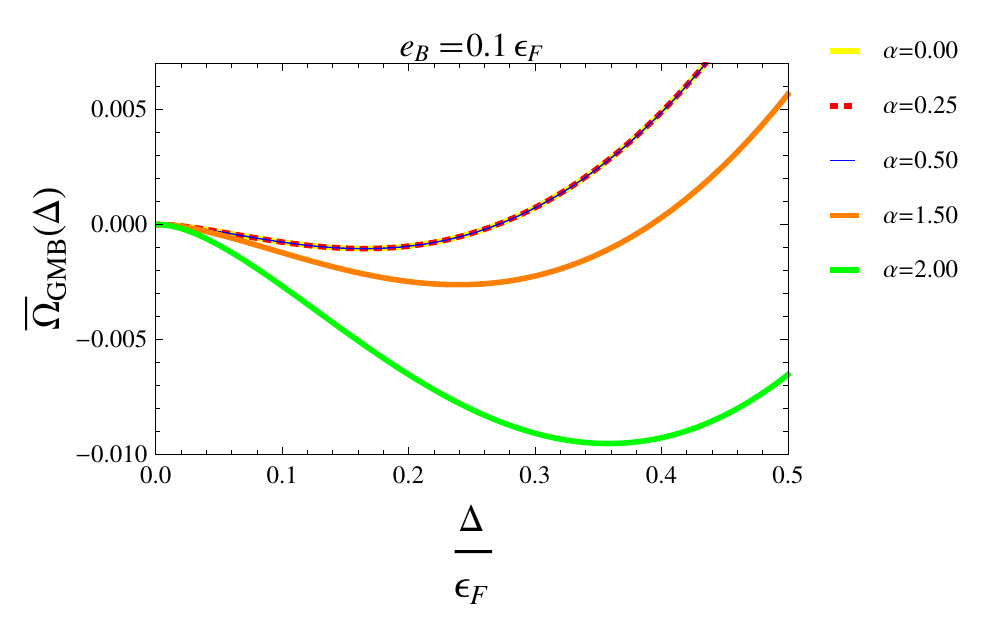}
\caption[]{(Color online) GMB corrected thermodynamic potential for different values of $\alpha=\frac{\lambda}{v_F}$ in the regime 
of $\alpha\ll1$ and $\alpha\gg1$. The parameters used are $\mu=1.0\epsilon_F$ and $T=0.01\epsilon_F$.}
\label{fig6}
\end{figure}

Here we show the influence of the GMB correction in the thermodynamic potential. Similarly to our analytical results, we separate our analysis in two 
limits: weak SOC and strong SOC. We identify these regions in our quantities by the indices $\rm{w}$ and $\rm{s}$ for weak SOC and strong SOC, 
respectively. In Fig.~\ref{fig6} we show the renormalized (dimensionless) GMB corrected thermodynamic 
potential $\bar{\Omega}\equiv\frac{\Omega\left(\Delta\right)-\Omega\left(0\right)}{k_F^2\epsilon_F}$ for different values of $\alpha$. As one can see, 
the minimum $\Delta$ increases with increasing $\alpha$ in both limits: GMBw and GMBs. Figure~\ref{fig6} should be compared with fig.~\ref{fig2} from 
which we can observe that the effect of the GMB correction is to decrease the MF ($\Delta_0(\alpha)$) minimum of the thermodynamic potential for a given $\alpha$.

\subsection{Critical Temperatures}

\subsubsection {Quasi-2D system}
The MF corrected transition temperature $T_c$ of a quasi-2D system is obtained by setting $\Delta=0$ in Eq.~(\ref{gapT}),

\begin{equation}
\int \frac{d^2 k}{(2 \pi)^2} \left[ \sum_{s=\pm} \left( \frac{1}{4 E_{\textbf{k},s} } \left(1-2 f_k(E_{\textbf{k},s} )\right) \right)-\frac{1}{2\epsilon_k
+|\epsilon_B|} \right] - N(0) F(\alpha)=0,
\label{gapT2}
\end{equation}
which shall be solved self-consistently with Eq.~(\ref{NumbEq}). In Eq.~(\ref{gapT2}), $F(\alpha)=1$ for $\alpha \ll 1$, and is 
given by Eq.~(\ref{Falpha}) for $\alpha \gg 1$. We also have that $ F(\alpha \to \infty)=1$. Note that we have used Eq.~(\ref{reg}) 
to regulate the ultraviolet divergence in the above equation, associated with the zero temperature logarithmically divergent term in 
Eq.~(\ref{gapT}). The difference between the equation above and the usual mean-field thermal gap equation is that the particle-hole 
fluctuation has been taken into account through the effective {\it s}-wave interaction $U_{eff}$.

\begin{figure}[htb]
\vspace{0.75cm} 
\includegraphics[width=10.0cm, height=6.5cm]{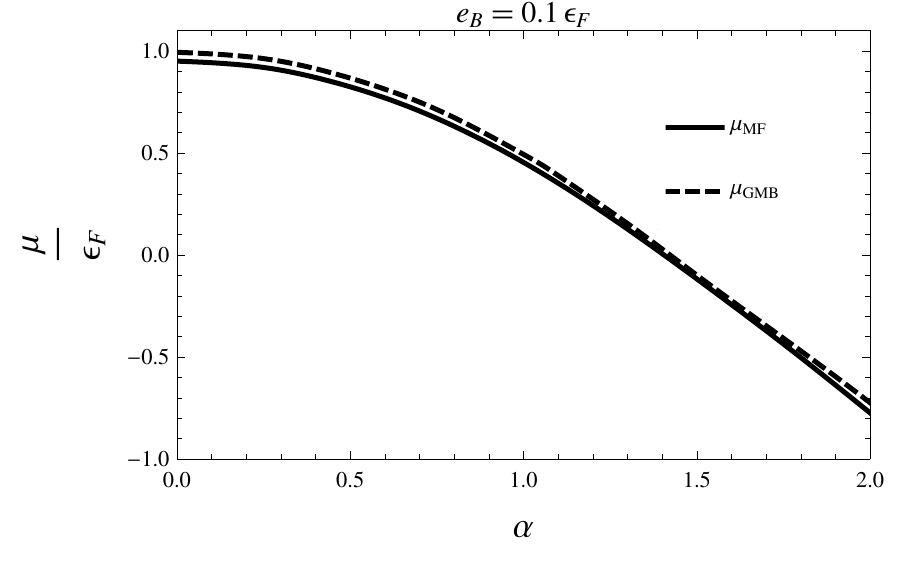}
\caption[]{Chemical potential in MF and GMB approximations for different values of $\alpha=\frac{\lambda}{v_F}$. We not differentiate the behaviors of the GMBw and GMBs approximations because they are practically the same. In this plot we use $T=0.01\epsilon_F$.}
\label{fig7}
\end{figure}

\begin{figure}[htb]
  \vspace{0.75cm} 
  \includegraphics[width=10.0cm, height=6.5cm]{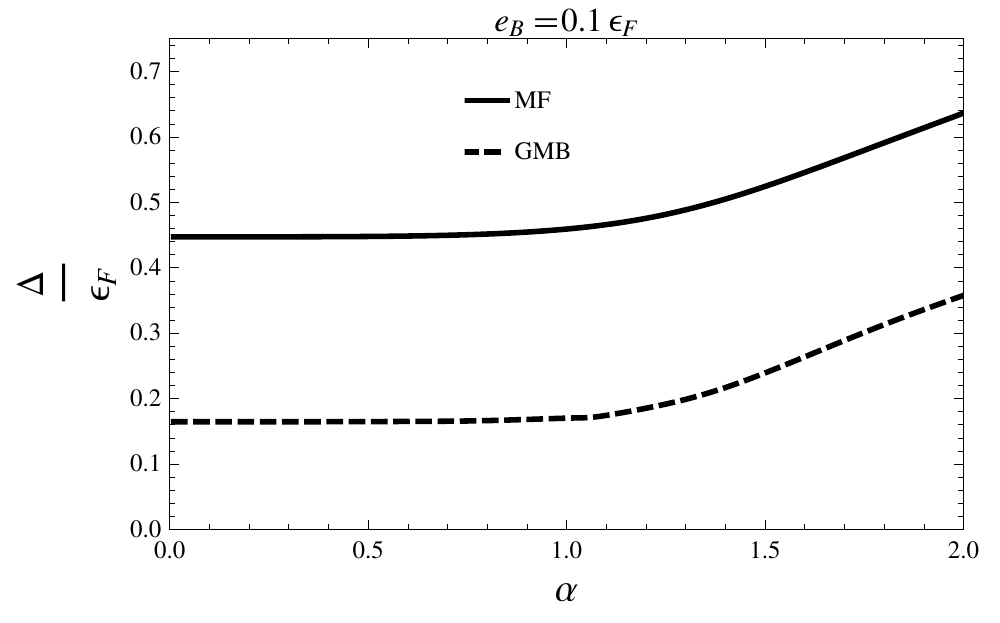}
\caption[]{Dimensionless superfluid pairing gap $\frac{\Delta}{\Delta_0}$ in MF and GMB approximations for different values of $\alpha=\frac{\lambda}{v_F}$ in regimes of $\alpha\ll1$ and $\alpha\gg1$. In this 
plot we perform an extrapolation of the GMBw results to GMBs. We use $T=0.01\epsilon_F$}
\label{fig8}
\end{figure}

\begin{figure}[htb]
  \vspace{0.75cm} 
  \includegraphics[width=10.0cm, height=6.5cm]{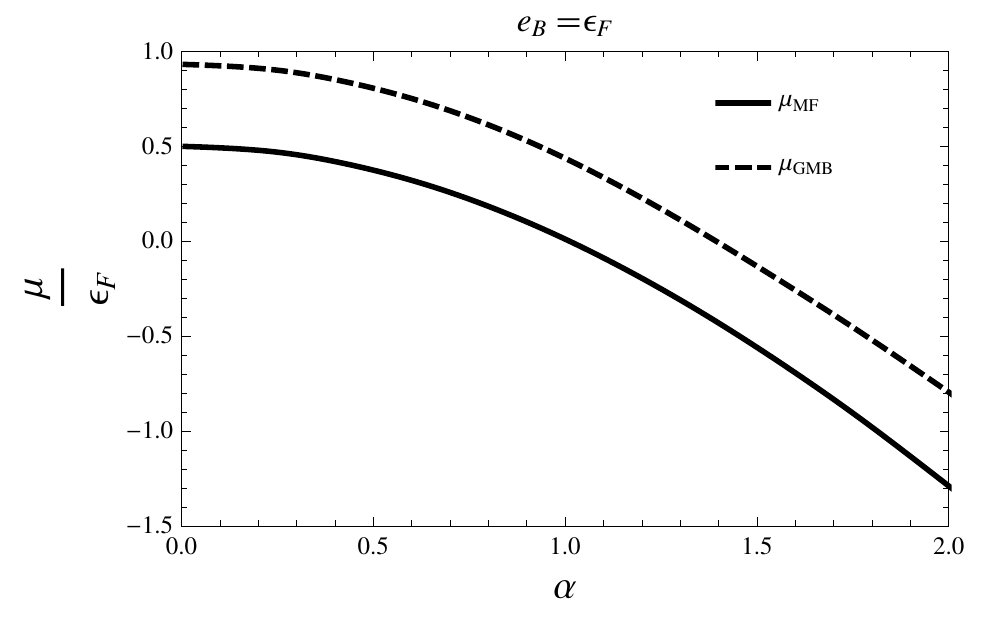}
\caption[]{Chemical potential in MF and GMB approximations for different values of $\alpha=\frac{\lambda}{v_F}$. We do not differentiate the behaviors of the GMBw and GMBs approximations since they are 
practically the same. In this plot we use $T=0.01\epsilon_F$.}
\label{fig9}
\end{figure}

\begin{figure}[htb]
  \vspace{0.75cm} 
  \includegraphics[width=10.0cm, height=6.5cm]{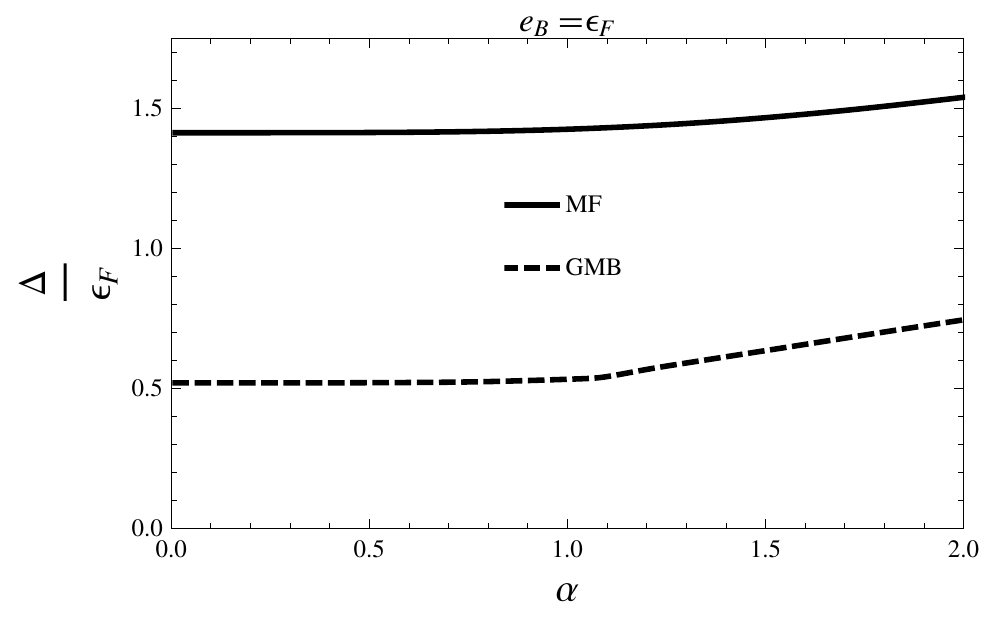}
\caption[]{Dimensionless superfluid pairing gap $\frac{\Delta}{\Delta_0}$ in MF and GMBs approximations for different values of $\alpha=\frac{\lambda}{v_F}$. In this plot we use $T=0.01\epsilon_F$}
\label{fig10}
\end{figure}

We solve numerically the system of gap and number equations self-consistently to find the superfluid pairing gap $\Delta$ and chemical potential $\mu$, for different values of the binding energy $e_B$. Motivated by recent MF results \cite{Chen} we perform a detailed investigation of the role of the strength of Rashba 
SOC coupling in the BCS-BEC crossover beyond MF approximation, i.e., considering the effect of induced interactions. 
For small binding energy (compared with Fermi energy $\epsilon_F$), $e_B=0.1\epsilon_F$,   Fig.~\ref{fig7} shows that the GMB correction does not change the value of the chemical potential and the BCS-BEC crossover is achieved for the same value of the  SOC strentgh given by the MF results. However,  Fig.~\ref{fig8} shows a considerable supression of $\Delta$ in the GMBs regime, compared with the MF result. If we take a larger value of the binding energy, $e_B=1.0 \epsilon_F$ for instance, the results of Fig.~\ref{fig9} show that the GMB correction increases the chemical potential by approximately $50\%$ and the crossover BCS-BEC happens for a larger value of SOC coupling when compared to the MF results. In Fig.\ref{fig10} we  see that the superfluid gap is enhanced in the region of strong SOC, as shown in Ref.~\cite{Chen} in the MF approximation. We show that considering the GMB correction, the MF enhancement of $\Delta$ is (significantly) suppressed in the region of weak (strong) SOC.

\begin{figure}[htb]
  \vspace{0.75cm} 
  \includegraphics[width=7.5cm, height=5.5cm]{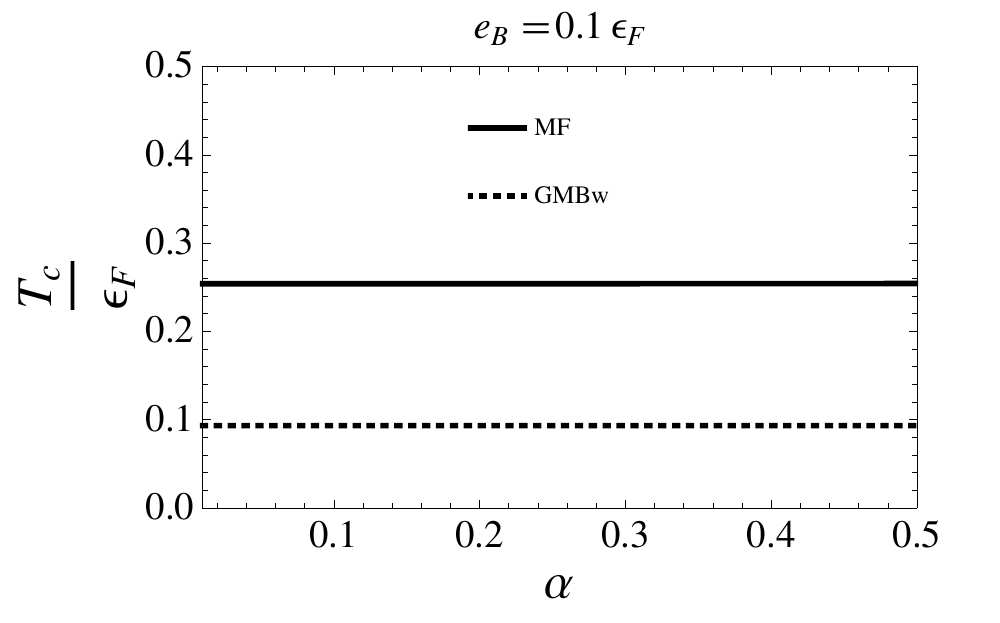}
  \includegraphics[width=7.5cm, height=5.5cm]{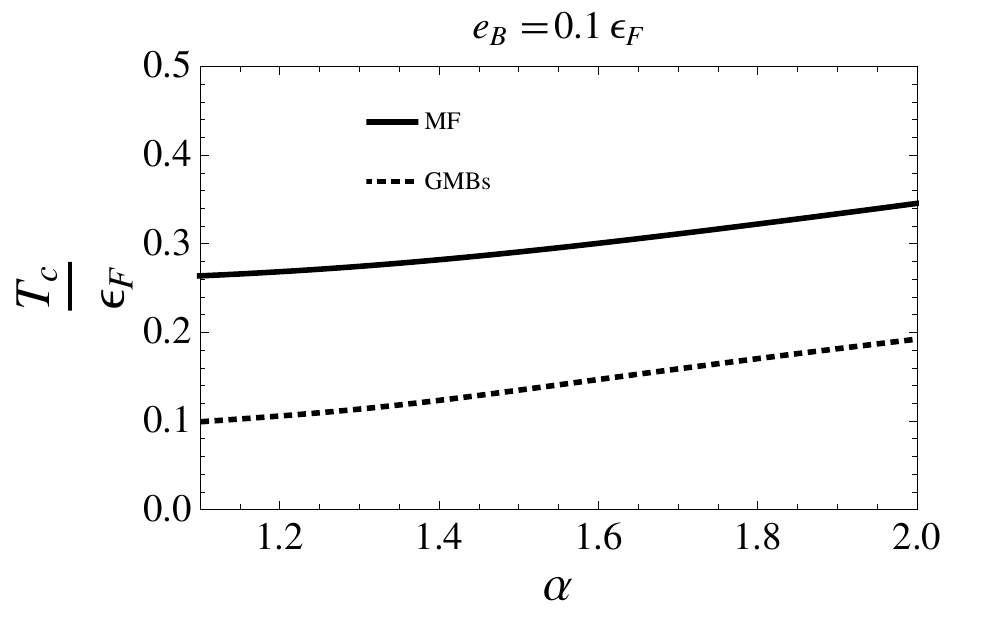}
\caption[]{Critical temperature in MF, GMBw and GMBs approximations for different values of $\alpha=\frac{\lambda}{v_F}$ in the regimes 
of $\alpha\ll1$ and $\alpha\gg1$. }
\label{fig11}
\end{figure}

\begin{figure}[htb]
  \vspace{0.75cm} 
  \includegraphics[width=7.5cm, height=5.5cm]{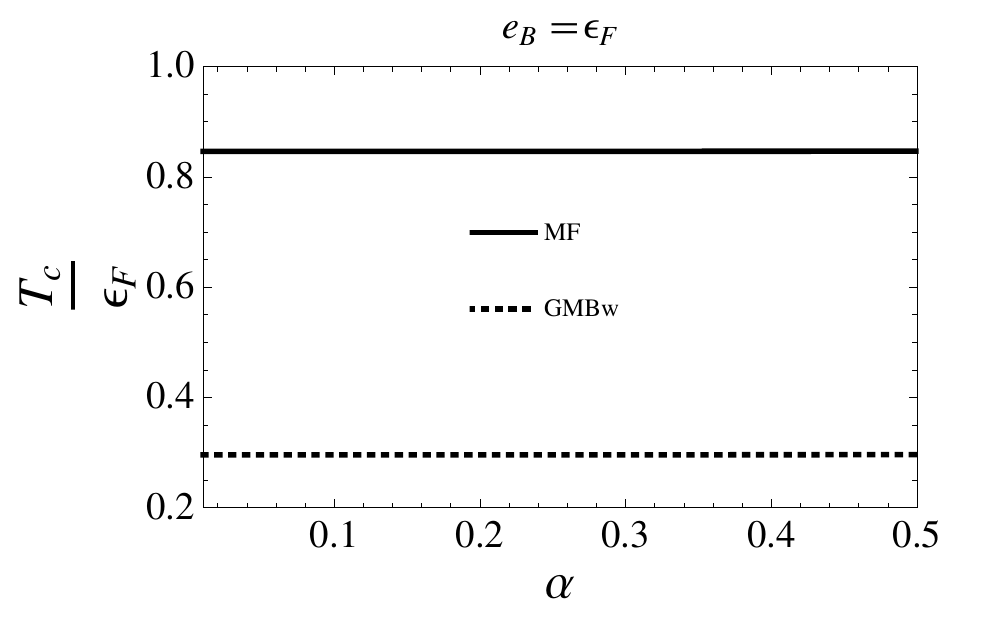}
  \includegraphics[width=7.5cm, height=5.5cm]{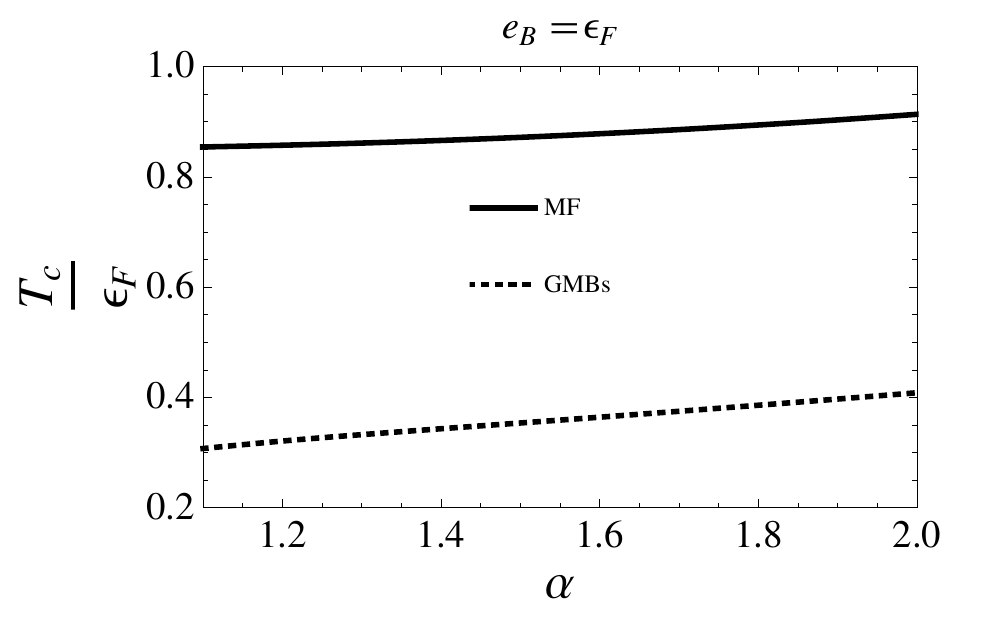}
\caption[]{Left panel: critical temperatures in MF and GMBw approximations in the regime of $\alpha\ll1$. Right panel: critical temperatures in MF 
and GMBs approximations in the regime of 
$\alpha\gg1$.}
\label{fig12}
\end{figure}

The critical temperature $T_c$ was obtained solving numerically the system of gap and number equations for $\Delta=0$. The results for both 
approximations are in Fig.~(\ref{fig11}) and Fig.~(\ref{fig12}).

\begin{figure}[htb]
  \vspace{0.75cm} 
  \includegraphics[width=6.5cm, height=5.5cm]{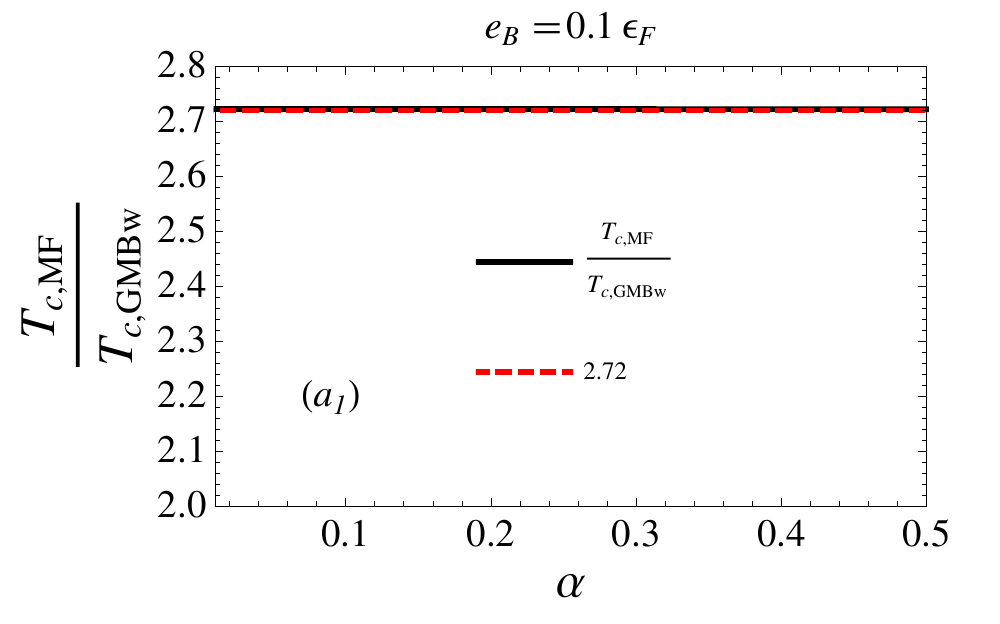}
  \includegraphics[width=6.5cm, height=5.5cm]{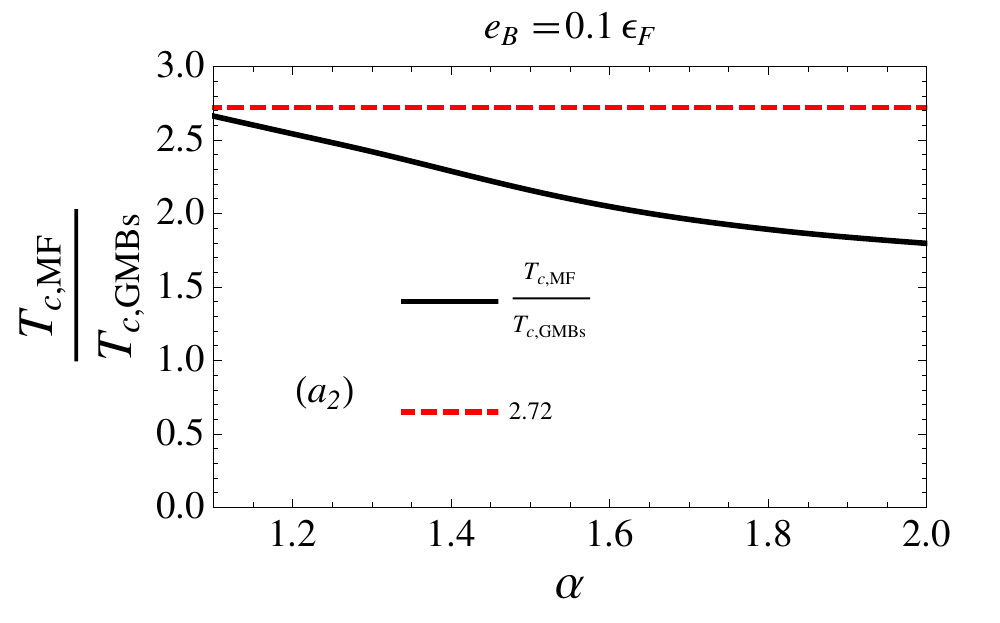}
  \includegraphics[width=6.5cm, height=5.5cm]{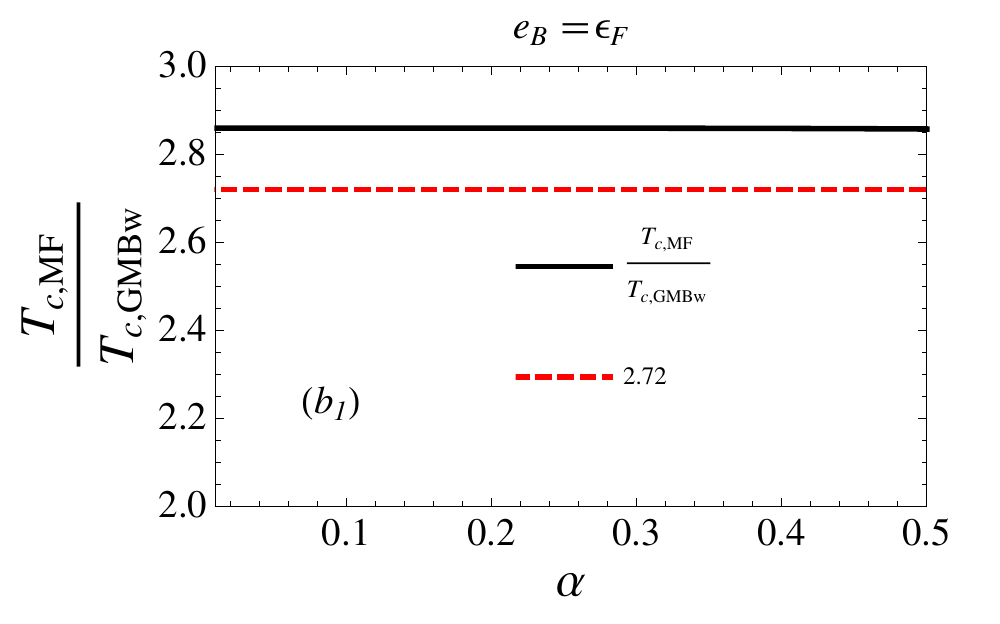}
  \includegraphics[width=6.5cm, height=5.5cm]{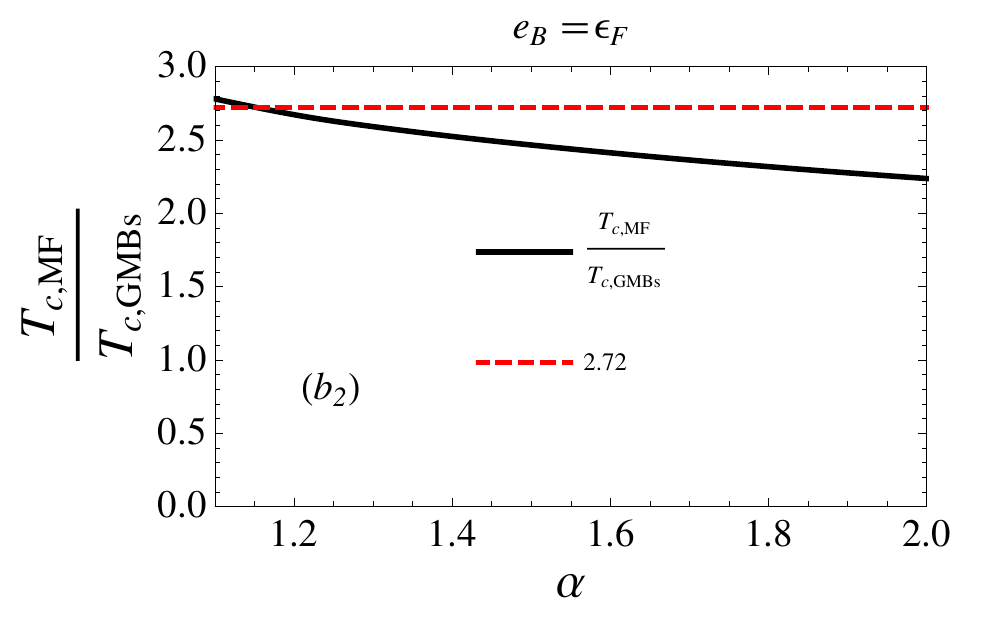}
\caption[]{(Color online) Suppression the MF critical temperature due the GMB corrections: ($a_1$) GMBw with $e_B=0.1\epsilon_F$, ($a_2$) GMBs with $e_B=0.1\epsilon_F$, ($b_1$) GMBw with $e_B=1.0\epsilon_F$ and ($b_2$) GMBs with $e_B=1.0\epsilon_F$. The value $2.72$ is the suppression of $T_c$ without SOC.}
\label{fig13}
\end{figure}

The suppression in the MF critical temperature due the GMB corrections are presented in Fig.\ref{fig13} for 
GMBw and GMBs corrections. We find that these critical temperatures have a similar trend as the SOC increases for both regimes of weak and strong SOC.

\subsubsection {Strictly 2D system}

In pure 2D systems, the relevant transition temperature is the Berezinskii-Kosterlitz-Thouless (BKT) transition~\cite{BKT} to a phase with 
quasi-long-range order. Below $T_{BKT}$ pairs of vortices and antivortices emerge, and will eventually condense as the temperature is 
lowered~\cite{Sa,Duan2,Devreese}. It has been shown very recently~\cite{Gong} that $T_{BKT}$ ``surprisingly'' decreases with increasing $\alpha$.

An asymptotic expression for the BKT temperature may be obtained~\cite{Loktev}

\begin{equation}
T_{BKT}=T_c -4\frac{T_c^2}{\epsilon_F}.
\label{TBKT}
\end{equation}
Our results agree with Ref.~\cite{Gong}, where $T_{BKT} \approx T_c$ for small $T_c$ (small $\alpha$) and $T_{BKT}$ decreases with increasing 
$T_c$ (increasing $\alpha$). Equation~(\ref{TBKT}) can also be used to obtain a GMB corrected BKT transition temperature $T_{BKT}^{GMB}$ as 
a function of $T_{c}^{GMB}$.

\subsubsection {Realistic SOC}

Notice that in practice, usual laser schemes for neutral atoms give rise to additional interactions besides the Rashba SOC. These include Dresselhaus and Zeeman terms~\cite{spielman} which are one-body terms that can be easily incorporated in the present work. However, more important is that the Rashba SOC is the dominant interaction in an expansion in terms of the laser coupling. The Dresselhaus and Zeeman terms can be controlled and the former made sufficiently small using strong laser intensities~\cite{spielman}.


\section{Conclusions}
\label{Conc}

We have studied the effects of induced interactions on the superfluid order parameter $\Delta$, chemical potential $\mu$ and transition temperature $T_c$ 
of a quasi-2D Fermi gas of atoms with Rashba SOC. As verified before in previous studies, the MF gap $\Delta$ and critical temperature $T_c$ increase with increasing (the strength 
of the SOC) $\alpha$, while $\mu$ decreases with increasing $\alpha$, signaling that the BCS-BEC crossover can occur for smaller values of the binding energy $e_B$, when compared to the case without SOC.
We find that taking into account the GMB correction there is a suppression of the MF enhancement of $\Delta$, 
and $T_c$ is reduced by a factor of $\approx 2.72$ for $\alpha \ll 1$ and by smaller factors when $\alpha  \gg 1$ (see Fig.~\ref{fig13}). This is a clear indication that the attractive interaction between particles becomes weaker due to the polarization of the medium, regardless external fields (lasers) acting on the atoms spins. 

Regarding the GMB corrected $\mu$, our calculations show that the BCS-BEC crossover takes place only for a value of $\alpha$ about $50 \%$ higher than the MF prediction (see Fig.~\ref{fig9}).

In conclusion, we have found that the superfluid parameters of a quasi-2D Fermi gas of atoms with SOC are considerably modified when the effects of induced 
interactions are properly taking into account. Our results are promising for achieving the superfluid transition in the regime of BCS pairing as well as the 
BCS-BEC crossover in current experiments.

\section{Acknowledgments}

This work was partially supported by CAPES, CNPq, FAPERJ,  and FAPEMIG (Brazilian Agencies). 



\end{document}